\documentclass[10pt,letterpaper]{article}
\usepackage{opex3}
\usepackage{epstopdf}
\usepackage{amssymb}
\usepackage{soul, color}
\usepackage{ulem}
\usepackage{cite}

\newcommand{\ket}[1]{\vert#1\rangle}

\begin{document}

\title{{
Efficient Bell state analyzer for time-bin qubits with fast-recovery WSi superconducting single photon detectors}}

\author{R.~Valivarthi,$^{1,2}$ I.~Lucio-Martinez,$^{1,2}$ A.~Rubenok,$^{1,2,6}$ P.~Chan,$^{1,3}$ F.~Marsili,$^{5}$ V.~B.~Verma,$^{4}$ M.~D.~Shaw,$^{5}$ J.~A.~Stern,$^{5}$ J.~A.~Slater,$^{1,2,7}$ D.~Oblak,$^{1,2}$ S.~W.~Nam$^{4}$ and W.~Tittel$^{1,2}$}

\address{$^{1}$ Institute for Quantum Science and Technology, University of Calgary, Calgary, AB, T2N 1N4, Canada \\
$^{2}$ Department of Physics and Astronomy, University of Calgary, Calgary, AB, T2N 1N4, Canada \\ 
$^{3}$ Department of Electrical and Computer Engineering, University of Calgary, Calgary, AB, T2N 1N4, Canada \\
$^{4}$ National Institute of Standards and Technology, Boulder, CO, 80305, USA\\
$^{5}$ Jet Propulsion Laboratory, California Institute of Technology, Pasadena, CA, 91109, USA \\
$^{6}$  Present Address: School of Physics, HH Wills Physics Laboratory, University of Bristol, Tyndall Avenue, Bristol, BS8 1TL, United Kingdom \\
$^{7}$  Present Address: Vienna Center for Quantum Science and Technology (VCQ), Faculty of Physics, University of Vienna, Boltzmanngasse 5, A-1090, Vienna, Austria}
\email{vrrvaliv@ucalgary.ca} 



\begin{abstract}
We experimentally demonstrate a high-efficiency Bell state measurement for time-bin qubits that employs two superconducting nanowire single-photon detectors with short dead-times, allowing projections onto two Bell states, $\ket{\psi^-}$ and $\ket{\psi^+}$. Compared to previous implementations for time-bin qubits, this yields an increase in the efficiency of Bell state analysis by a factor of thirty.
\end{abstract}

\ocis{(270.5565) Quantum communications, (270.5568) Quantum cryptography, (270.5570) Quantum detectors}




\section{Introduction}
Bell state measurements (BSMs) play a key role in linear optics quantum computation and many quantum communication protocols, e.g. quantum repeaters \cite{Sanguard11}, quantum teleportation~\cite{Bennett93}, dense coding \cite{Mattle96} and some quantum key distribution protocols \cite{Lo2012}. A complete BSM allows projecting any two-photon state deterministically and unambiguously onto the set of four maximally-entangled Bell states, i.e. 

\[
\ket{\phi^\pm} = \frac{1}{\sqrt{2}}(\ket{00} \pm \ket{11})
\]
and
\[
 \ket{\psi^\pm}= \frac{1}{\sqrt{2}}(\ket{01} \pm \ket{10}).
\]

\begin{figure}[h]
\centering
\includegraphics[width=\textwidth]{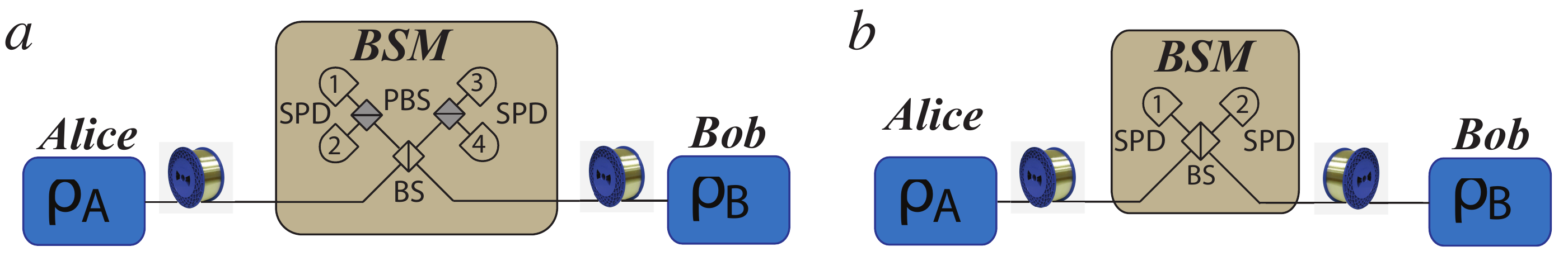}
\caption{Experimental setup used to perform BSMs for a) polarization qubits and b) time-bin qubits. Density matrices $\rho_A$ and $\rho_B$ characterize the states of the photons emitted at Alice's and Bob's, respectively. Optical components: beam splitter (BS) and single photon detectors (SPD).}
\label{fig:BSM}
\end{figure}

\noindent Unfortunately, it has been shown that a complete BSM is impossible when using linear optics and no auxiliary photons: the probability for a BSM to succeed (henceforward referred to as efficiency, $\eta_{BSM}$) in the case of two photons in completely mixed input states (e.g. two photons that are members of different entangled pairs)  is, in principle, limited to 50\% \cite{Lutkenhaus99}. The standard approach to Bell state analysis uses a 50/50 beam splitter followed by single-photon detectors that allow (possibly using additional external optical elements) discriminating between orthogonal qubit states $\ket{0}$ and $\ket{1}$ (see Fig. \ref{fig:BSM}). This approach allows one to unambiguously project onto $\ket{\psi^-}$ and $\ket{\psi^+}$. 

For instance, when implementing a BSM for polarization qubits, a projection onto $\ket{\psi^-}$ occurs if the two photons exit the beam splitter through two different ports and are detected in orthogonal polarizations, leading to detections in detectors 1 and 4, or detectors 2 and 3 (for an illustration see Fig. \ref{fig:BSM}(a) ). Furthermore, projections onto $\ket{\psi^+}$ happen if the two photons exit the beam splitter through the same port and, as before, are detected in orthogonal polarization states. This leads to detections in detectors 1 and 2, or detectors 3 and 4 (see Fig. \ref{fig:BSM}(a)). Other coincidence detections correspond to projections onto product states $\ket{H}\ket{H}\equiv\ket{0}\ket{0}$ and $\ket{V}\ket{V}\equiv\ket{1}\ket{1}$. Hence, this scheme allows achieving the maximum efficiency value of 50\% if one considers single photon detectors with unity detection efficiency. Assuming realistic detectors with efficiency $\eta_{det}$, the BSM efficiency is reduced to

\begin{equation}
\eta_{BSM}=\frac{1}{2}\eta_{det}^2.
\label{eq:efficiency}
\end{equation}

\begin{figure}[h]
\centering
\includegraphics[width=\textwidth]{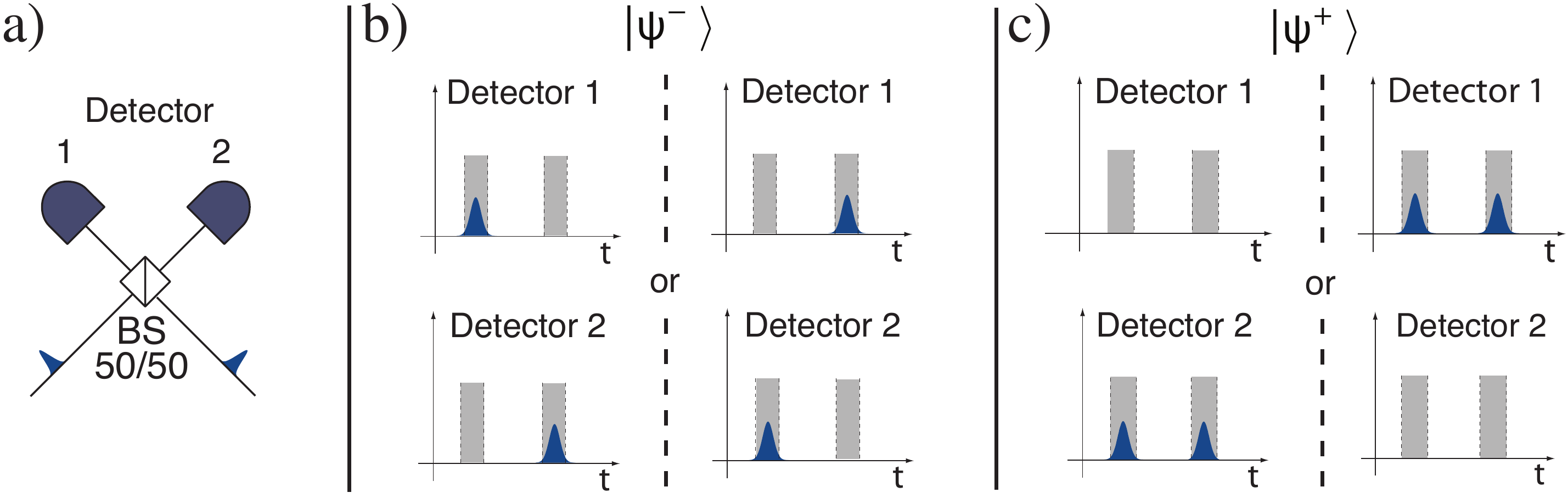}
\caption{a) General setup for Bell state measurement for time-bin qubits using linear optics and single photon detectors (SPD). b) Detection pattern for projections onto $\ket{\psi^-}$ and (c) $\ket{\psi^+}$.}
\label{fig:patterns}
\end{figure}

\noindent
In addition to polarization, another widely used degree of freedom to encode qubits is time. In this case photons are generated in a superposition of two temporal modes $\ket{early}\equiv\ket{0}$ and $\ket{late}\equiv\ket{1}$ -- so-called time-bin qubits.   
Time-bin qubits are particularly well suited for transmission over optical fiber (and thus generally encoded into photons at telecommunication wavelength), and have been used for a large number of experiments \cite{plug&play,Brendel,Tittel2000}, including experiments that require projections onto Bell states \cite{Marcikic2003, DeRiedmatten2005,Jin2013,Rubenok13,Liu13}. 
BSMs with time-bin qubits generalize the scheme introduced above for polarization qubits but require only a single beam splitter as illustrated in Fig.~\ref{fig:BSM}(b).
The temporal detection pattern of photons after passing the beam-splitter (see Fig.~\ref{fig:patterns}(a)) corresponds to different bell-state projections.
A projection onto the singlet $\ket{\psi^-}$ state occurs if one of the two detectors registers a photon in the early time bin and the second detector registers a photon in the late time bin (see Fig. \ref{fig:patterns}(b)). On the other hand, a projection onto $\ket{\psi^+}$ happens if a detector registers one photon in the early time bin, and the same detector detects the second photon in the late bin (see Fig. \ref{fig:patterns}(c)). 

However, a problem arises if the detection of a photon is followed by dead-time during which the detector cannot detect a subsequent photon. For example, for commercial InGaAs-based single photon detectors (SPDs), which are widely used for quantum communication applications including BSM with time-bin qubits, this dead-time is typically around 10~$\mu$s (to the best of our knowledge, the exceptions are \cite{Liu13}, where frequency conversion and Si-APDs were employed, and \cite{Dixon09, Zhang09}, where InGaAs-based SPDs with dead-times of 2~ns and 10~ns and quantum detection efficiencies of $\approx$10\% have been reported. However, none of the last-mentioned detectors have been used for BSMs with time-bin qubits.).
This dead-time is necessary to suppress afterpulsing due to trapped carriers that are released after a detection and cause subsequent detection signals \cite{Zhang08}. 
The dead-time of the detectors previously employed for the BSM have always been orders of magnitude longer than the maximally achievable time difference between early and late temporal modes, which is limited by either the ability to phase-stabilize the required widely unbalanced Mach-Zehnder interferometers (the path-length difference being proportional to the time-bin separation) or by the required frequency stability of the source lasers used for generating the time-bin qubits.
Thus, commercial InGaAs SPDs have usually restricted BSMs with time-bin qubits to projections onto $\ket{\psi^-}$, reducing the maximum efficiency of the BSM from 50\% to 25\%. The only exception is \cite{Houwelingen06}, where the unambiguous projections onto three Bell states with theoretically maximum probability of 5/16$\approx$31\% was proposed and a proof-of-principle demonstration reported. Taking a typical detection efficiency for InGaAs SPDs of 15\% into account, the highest efficiency of a BSM for time-bin qubits is currently thus only around 1\%. This includes the demonstrations reported in \cite{Liu13, Dixon09, Zhang09}  and \cite{Houwelingen06}.

In this paper we present an efficient BSM for time-bin qubits encoded into telecommunication photons with projections onto the $\ket{\psi^-}$ as well as the $\ket{\psi^+}$ Bell state. Towards this end, we employ two superconducting nanowire single photons detectors (SNSPDs), which, in addition to short dead-times, feature low dark count rates and system detection efficiencies of 76\%. This leads to an increase of $\eta_{BSM}$ by a factor of thirty compared to previous implementations, which is an important improvement in view of future applications of quantum information processing involving many BSMs, e.g. quantum repeaters.

The remainder of this article is structured as follows. In section \ref{sec:det} we describe the single-photon detectors employed to perform the measurements, and in section \ref{sec:experiment} we present the details of the experimental setup. The results of our measurements are presented and discussed in section \ref{sec:results}. Finally, in section \ref{sec:conclusions}, we present our conclusions and outlook.

\section{Superconducting single photon detectors with short dead-times}\label{sec:det}

Recent years have seen great progress in the development of single-photon detectors for telecommunication wavelengths. Arguably, the best detectors today are based on the transition of a superconducting nanowire into the resistive state~\cite{Natarajan2012a}, and many benchmark results have been reported with these SNSPDs. This includes dead-times as small as 10~ns \cite{Kerman2006a,Robinson2006a}, and quantum efficiencies up to 93\% at 1550~nm \cite{Marsili13}. Furthermore, unlike InGaAs SPDs, which require gating, SNSPDs are inherently free running, show no afterpulsing, and feature very low dark count rates on the Hz level \cite{Marsili13}.

\begin{figure}[h]
\centering
\includegraphics[width=\textwidth]{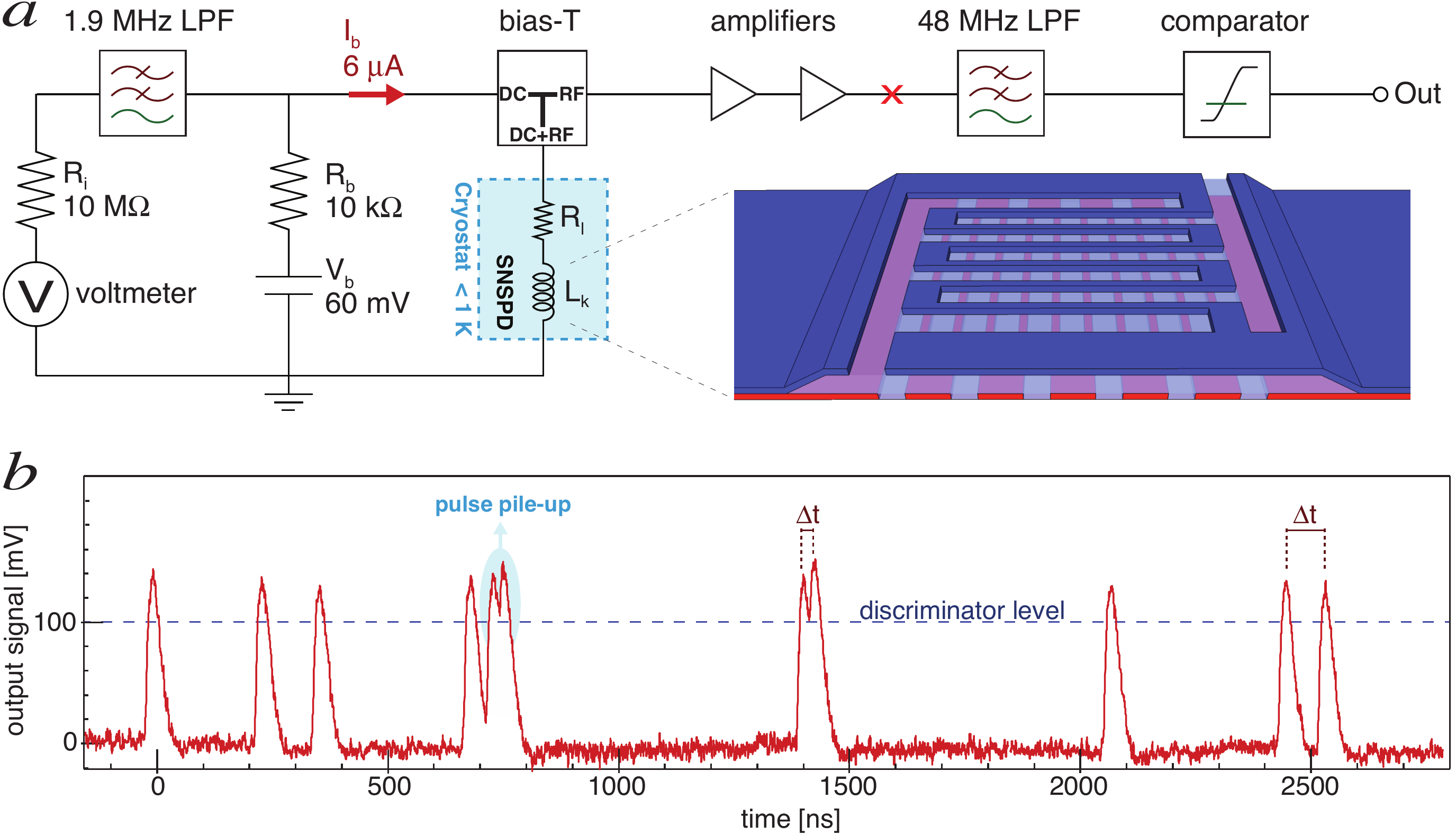}
\caption{Detector setup and signal. a) Electrical diagram of the SNSPD setup. The $R_b=10~\mathrm{k}\Omega$ bias resistor translates the 60~mV bias voltage into a $I_b=6~\mu$A bias current, which is directed to the superconducting detectors via the DC-port of the bias-T. The RF-port of the bias-T directs the photon detection signal through two amplifiers and a low-pass-filter (LPF) to a comparator, which generates a TTL output signal. The parallel connected voltmeter measures the voltage drop over the SNSPD and allows verifying that it is in the superconducting state. The panel also shows a sketch of an SNSPD consisting of two meanders. b) Single photon detection signals of detector 2 immediately after the amplifiers (marked by an x in Fig. a). A few detection inter-arrival times $\Delta t$ are indicated for illustration.}
\label{snspd:fig1}
\end{figure}

We employ SNSPDs that have been developed and fabricated at the National Institute for Standards and Technology (NIST) and the Jet Propulsion Laboratory (JPL). The detectors are based on one, or two mutually orthogonal, tungsten silicide (WSi) nanowire meanders (we refer to the two different detectors as detector 1 and 2, respectively -- see Fig~\ref{snspd:fig1}(a) for a sketch of detector 2. The detector with two meanders features a detection efficiency that is highly insensitive to photon polarization \cite{Verma12}, whereas the single meander version experiences up to 10\% variation in efficiency at different polarizations. The two SNSPDs are mounted on an adiabatic diamagnetic refrigeration (ADR) stage inside a pulse-tube cooler, and are operated at a temperature around 800~mK. The setup for characterizing and operating the detectors is sketched in Fig.~\ref{snspd:fig1}(a). The SNSPDs are represented by a kinetic inductance $L_k$ and load resistance $R_l=50~\Omega+R_s$, where the first term on the right-hand-side is the impedance of the output coaxial cable and $R_s$ is an additional and optional series resistor. 
A sample of the detection signal is shown in Fig.~\ref{snspd:fig1}(b).
The detector quantum efficiencies were measured at 1550~nm wavelength to be $77.5\pm0.7$\% and $76.2\pm0.9$\% for detectors 1 and 2, respectively.

\begin{figure}[h]
\centering
\includegraphics[width=\textwidth]{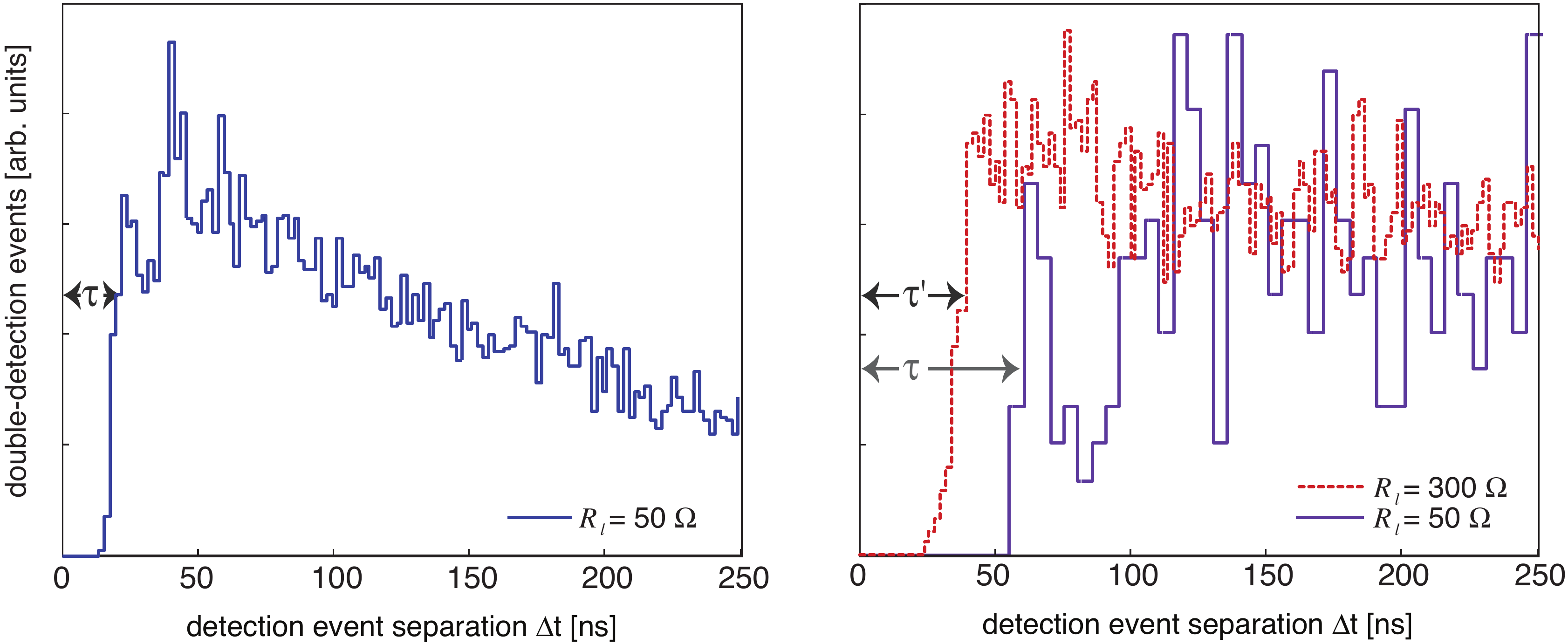}
\caption{Detection dead-times. Histograms of detection inter-arrival times for SNSPD 1 and 2 in the left and right panel, respectively. Solid lines correspond to the setup with $R_l=50~\Omega$ (given by the impedance of the coaxial cable), while the dashed line shows the result when a $R_l=350~\Omega$ resistor is connected to detector 2 inside the cryostat. For $R_l$=50$~\Omega$ we find $\tau\approx$~30~ns for detector 1 and $\tau\approx$~100~ns for detector 2. The dead-time of detector 2 is reduced to around 40~ns when using $R_l$=350$~\Omega$.}
\label{snspd:fig3}
\end{figure}

To assess the detector dead-times, we illuminate the SNSPDs with weak continuous wave (cw) light and log the time $\Delta t$ between subsequent detections, as illustrated in Fig.~\ref{snspd:fig1}(b). Histograms of these  inter-arrival detection times reveal the minimum time separation $\tau$ between detection events -- during this time, the SNSPDs cannot detect another photon either because of the intrinsic time it takes the current to reflow or because of a pulse pile-up in which the signal does not cross the discriminator level between two consecutive incident photons and thus only the first detection event is registered. The measurements, with a $50~\Omega$ coaxial cable attached to the detectors, shown in Fig.~\ref{snspd:fig3} by the solid lines, gives a dead-time $\tau$ on the order of 30~ns for detector 1 and 100~ns for detector 2. This dissimilarity of dead-time is due to the difference in kinetic inductance of the detectors \cite{Kerman2006a}. Hence, to allow projections onto the $\ket{\psi^+}$ state using detector 2, the time-bin separation would have to be on the order of 100~ns.

As argued above, it is desirable to reduce the SNSPD dead-time. Previous studies have shown $\tau\propto L_k/R_l$, and as the kinetic inductance is related to the inherent geometry and material properties of the SNSPD (which cannot be easily modified), we focus on increasing $R_l$ as a means of reducing $\tau$ \cite{Yang2007a}. To that end we put a $R_s=300~\Omega$ resistor in series with SNSPD detector 2. The resistors are regular ceramic surface-mount resistors and are connected to the SNSPDs after a 10~cm long coaxial cable. The resulting inter-arrival time statistics is plotted as a dashed line in Fig.~\ref{snspd:fig3}. We see that the new dead-time of detector 2, $\tau'$, is significantly reduced to around 40~ns. The discrepancy between the 7 fold increase in $R_l$ and the resulting 2.5 fold decrease in the dead-time is most likely due to uncertainty of the exact value of $R_s$ at low temperatures and limitations on our ability to discriminate subsequent detections due to pulse pile-up. One might conclude that an additional increase of the load resistance would further reduce the dead-time. However, we anticipate that with larger values of $R_l$ the detector would begin to latch (i.e. not return to the superconducting state after the detection of a photon).

\section{Experimental setup}\label{sec:experiment}
\begin{figure}[h!]
\centering
\includegraphics[width=\textwidth]{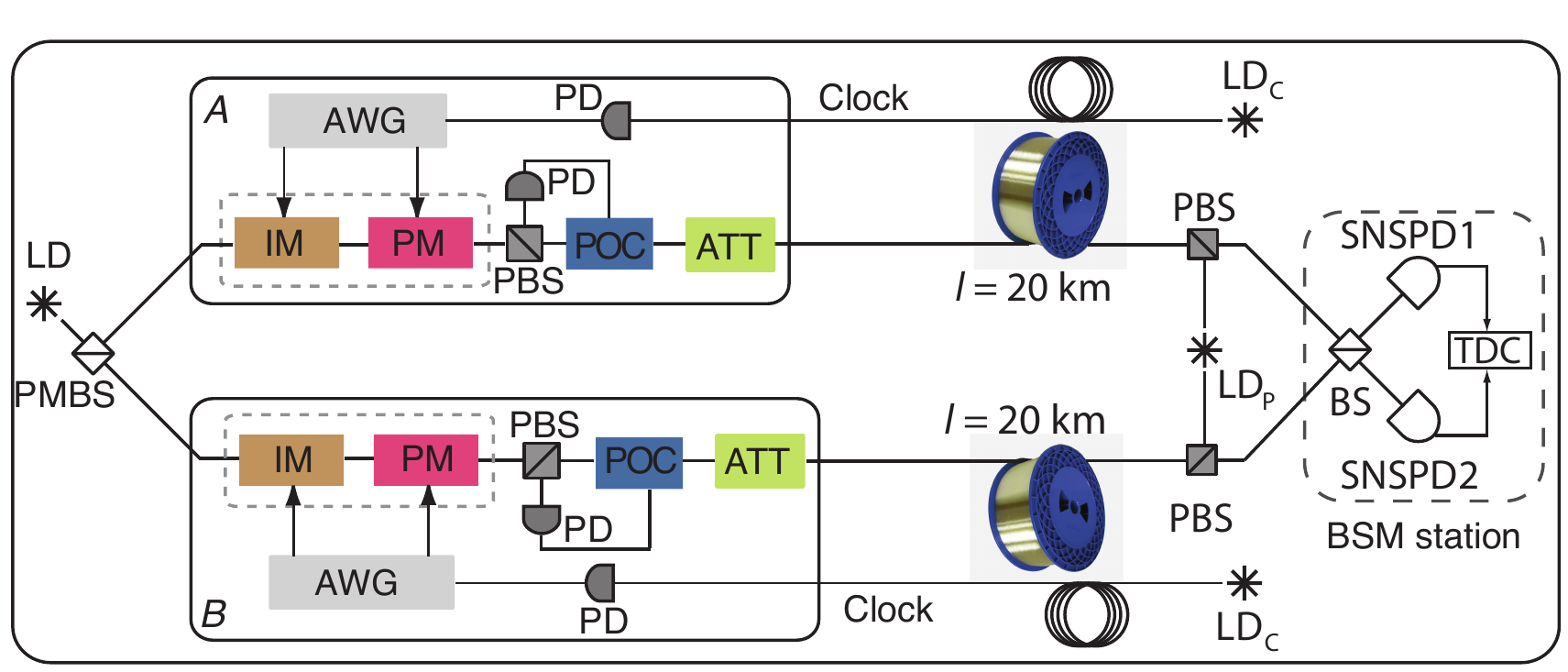}
\caption{Schematic of the experimental setup employed for a BSM with time-bin qubits. LD, laser diode; PMBS, polarization maintaining beam splitter; IM, intensity modulator; PM, phase modulator; PBS, polarization beam splitter; POC, polarization controller; PD, photodiode; BS, beam splitter; AWG, arbitrary waveform generator; ATT, variable optical attenuator; SNSPD, superconducting nanowire single-photon detector. The lasers LD$_C$ and LD$_P$ are used for timing and polarization feedback control, respectively, which is further explained in \cite{Rubenok13}.}
\label{fig:setup}
\end{figure}

Our experimental setup is similar to that described in \cite{Rubenok13}. As depicted in Fig.~\ref{fig:setup}, a stabilized cw laser emits polarized light at 1550 nm. The light is split by a polarization maintaining finer-optic beam splitter, and travels to two different stations, which we will refer to as Alice (A) and Bob (B). At each station, light is sent through intensity modulators that carve 0.5~ns long pulses, which, after appropriate attenuation, form time-bin qubit states encoded into laser pulses with mean photon number well below one. For instance, $\ket{0}$ corresponds to an attenuated laser pulse in an early temporal mode, $\ket{1}$ corresponds to a laser pulse in a late temporal mode, and $\ket{+} \equiv (\ket{0} + \ket{1})/\sqrt{2}$ is generated by opening the intensity modulator twice in a row, generating photons in a coherent superposition of  early and late temporal modes. The subsequent phase modulator allows applying a $\pi$ phase shift to the late temporal mode, which results in generating $\ket{-} \equiv (\ket{0}-\ket{1})/\sqrt{2}$. Qubits are created at a repetition rate of 5 MHz, and the two temporal modes are separated by 75 ns. Finally, each qubit (one generated at Alice's and one at Bob's) is sent through a polarization controller and 20 km of spooled fiber, which introduce random global phase shifts, and arrive at the Bell state analyzer where the BSM is performed using a beam splitter and two SNSPDs. Detection statistics is collected using a time-to-digital converter for various combinations of mean photon numbers per qubit generated at Alice's and Bob's, and is recorded on a PC. 

It is important to recall that, for a BSM, the two photons impinging on the beam splitter must be indistinguishable in all degrees of freedom: polarization, arrival time, and frequency. 
Frequency indistinguishability is particularly important when working with $\ket{\pm}$ time-bin qubit states, as a frequency difference $\Delta\nu$ translates into a difference $\Delta\phi$ between the phases characterizing the superposition of the two time-bin qubit states according to $\Delta\phi=2\pi\Delta\nu t_0$, where $t_0$ denotes the temporal separation between $\ket{0}$ and $\ket{1}$. While a constant phase difference (due to a constant frequency difference) can be compensated for during qubit preparation, having time varying phase differences becomes problematic once the variation of the phase difference exceeds a few degrees. Consequently, the time-bin separation is not only constrained by the dead-time of the detectors, but also by the frequency stability of the light sources (assuming independent sources). For example, for our time-bin separation of $t_0$=75 ns, the two lasers must be frequency stable at least within $\sim$185 kHz over the duration of  a measurement to keep the phase error under $5^o$. Unfortunately, lasers with such frequency stability are currently not commercially available. To circumvent this problem, we used only one laser in our experiment, which allowed Alice and Bob to generate time-bin qubits with stable phase relation. Finally, to ensure indistinguishability in polarization and arrival time, we implemented feedback control as described in \cite{Rubenok13}. 

\section{Results}\label{sec:results}
To characterize the reliability and efficiency of our Bell state analyzer, we work within the framework of the measurement-device-independent quantum key distribution (MDI-QKD) protocol \cite{Lo2012}. 
In MDI-QKD two parties, Alice and Bob, prepare qubits that are sent over channels to be projected onto entangled states via a BSM, thus establishing an entangled channel that allows for the generation of a correlated key.
Conversely, the possibility to generate highly correlated bits using an MDI-QKD type setup allows one to draw conclusions about the quality of the BSM.  To demonstrate efficient Bell state measurements with time-bin qubits, Alice and Bob prepare various combinations of qubit  states, encoded into attenuated laser pulses with one out of three possible mean photon numbers (0.11, 0.05 and 0) and with both qubits belonging to the same basis 
i.e. $\ket{\psi}_A,\ket{\psi}_B \in \{ \ket{0},\ket{1} \}$ or  $\ket{\psi}_A,\ket{\psi}_B \in \{ \ket{+},\ket{-} \}$
, and send them to the Bell state analyzer. We define the \textit{z}-basis to be spanned by $\ket{0}$ and $\ket{1}$, and the \textit{x}-basis to be spanned by $\ket{+}$ and $\ket{-}$. For each combination of states and mean photon numbers, we record the number of projections onto $\ket{\psi^+}$ and $\ket{\psi^-}$.

\subsection{Error rates}
An important criterion for assessing the possibility for BSMs with time-bin qubits are error rates, which, for each basis and Bell state, are given by the number of erroneous projections (e.g. projections onto $\ket{\psi^-}$ if the two input states were identical) divided by the total number of projections onto that Bell state. Towards this end, qubits should be encoded into true single photons. As we use attenuated laser pulses instead, which feature Poissonian-distributed photon numbers, we use a decoy state protocol \cite{Wang2013} to assess upper bounds $e_{11}$ for the error rates that we \textit{would} have measured \textit{had} we used true single photon inputs. (Here the subscripts $1$ refer to the single photon components of the Poissonian photon number distributions) These bounds are calculated from the Bell-state projections measurements using the three above mentioned mean photon numbers and the resulting error rates. The upper bound on the inferred single photon error rates are listed in table \ref{tab:single} below.
\begin{table}[h!]
\caption{Bounded error rates $e_{11}^z$ and $e_{11}^x$ for two single photon inputs (one at Alice's and one at Bob's) with both photons prepared in the $z$ and $x$ basis, respectively. The rates are extracted from the measured data using a decoy state method \cite{Wang2013}. }
\begin{center}
\begin{tabular}{|c|c|c|}
\hline
Error rates & Projections onto $\ket{\psi^-}$ & Projections onto $\ket{\psi^+}$\\
 &  (\%) & (\%) \\
\hline
$e_{11}^z$  & 0.44$\pm$0.07 & 0.80$\pm$0.07 \\
\hline
$e_{11}^x$  & 3.6$\pm$0.8 & 6.7$\pm$0.8 \\
\hline
\end{tabular}
\end{center}
\label{tab:single}
\end{table}%

The results are close to ideal, in particular regarding the error rate for the $z$-basis, which exceeds the ideal outcome of 0\% by only 0.44\% and 0.80\% for projections onto $\ket{\psi^-}$ and $\ket{\psi^+}$, respectively. This is a very good result, especially given that Alice and Bob are separated by 40 km of spooled fiber. 
The remaining errors are due to (almost negligible) background light leaking through Alice's and Bob's intensity modulators (featuring 50~dB extinction ratio) and detector dark counts (around 10 Hz, including detector counts due to blackbody radiation).  
For the $x$-basis, the error rates exceed the ideal outcome of 0\% by 3.6\% and 6.7\% for the $\ket{\psi^-}$ and $\ket{\psi^+}$ projections, respectively. 
We attribute the increment in the error rates compared to those of the z-basis to phase errors occurring during the preparation of the $\ket{-}$-state. In addition the gap between the bound on $e_{11}$ and its actual value is larger. This poorer performance of the decoy state analysis is due to errors in the raw data arising from multi-photon contributions (e.g. two photons arriving from Alice and one photon from Bob) \cite{Rubenok13}, which partially propagate into the calculated bound for $e_{11}^x$.

\subsection{Efficiency}
While error rates allow assessing if the BSM is functioning correctly, an equally important measure is the efficiency of the Bell state analyzer.  
As in the previous section, we use the decoy state protocol \cite{Wang2013}  to find a lower bound on  the number of projections onto $\ket{\psi^+}$ and $\ket{\psi^-}$ that originate from the emission of single photons at Alice's and Bob's. The number of such projections per clock cycle, $Q_{11}^{x,z}$ (where $x$, $z$ denotes the basis in which the qubits have been prepared), then allows us to calculate the BSM efficiency for each basis and Bell state using 

\begin{equation}
Q_{11}^{x,z} = P_1(\mu) P_1(\mu) t^2 \eta_{BSM}^{x,z}.
\end{equation}
Here, $P_1(\mu)$ refers to the probability of emission of a single photon per (Poissonian distributed) source, $t$ denotes to the transmission between Alice or Bob and the Bell state analyzer, and $\eta_{BSM}^{x,z}$ is the basis-dependent efficiency of the BSM. The results for $\eta_{BSM}$ are listed in table \ref{tab:results}.

\begin{table}[htdp]
\caption{Bell state measurement efficiencies extracted from measured data using a decoy state method \cite{Wang2013}.}
\begin{center}
\begin{tabular}{|c||c|c||c|}
\hline
 & Efficiency for projections & Efficiency for projections & Total efficiency (\%)\\
Basis & onto $\ket{\psi^-}$ (\%) & onto $\ket{\psi^+}$ (\%) & \\
\hline
$z$ & 13.6$\pm$0.2 & 14.5$\pm$0.2 & 28.1$\pm$0.3 \\
$x$ & 14.5$\pm$0.4 & 15.3$\pm$0.4 & 29.8$\pm$0.6 \\
\hline
\end{tabular}
\end{center}
\label{tab:results}
\end{table}

We note, first, that the values for the total efficiencies per basis differ by only $1.7\%$, confirming that we can perform all projections with almost equal probability. In particular, this shows that the detectors have indeed fully recovered after 75 ns. Second, we find that the efficiency averaged over the $x$, $y$ and $z$ bases (where we made the physically motivated assumption that the efficiency in the $y$-basis, which we did not measure, equals the one measured in the $x$-basis), $\eta_{BSM}$, corresponds to that estimated using Eq.\ref{eq:efficiency} and taking into account the measured detector quantum efficiencies:

\begin{eqnarray}
\eta_{BSM} &=& \frac{1}{3}\big (\eta_{bsm,z}+2\eta_{bsm,x}\big )=(29.3 \pm 0.4)\% \\
&\approx& \frac{1}{2}\eta_{det}^2=\big (29.5 \pm 0.4 \big )\%. \nonumber
\end{eqnarray}
\noindent 
Furthermore, we point out that the efficiency is a factor of $\approx$ 30 higher than what has previously been obtained with time-bin qubits. Finally, we note that our average BSM efficiency is only 2.3\% below the theoretical maximum of 5/16$\approx$31\% (assuming detectors with 100\% efficiency) achievable with previously implemented schemes \cite{Houwelingen06}.  

\section{Conclusions and Outlook}\label{sec:conclusions}
We have described and demonstrated how to perform efficient Bell state analysis with time-bin qubits using linear optics and no additional photons. By employing SNSPDs with short dead-times, it is possible to project not only onto the $\ket{\psi^-}$, but also onto the $\ket{\psi^+}$ Bell state. Together with the high quantum efficiency of the SNSPDs, this improved the efficiency of Bell state measurements with time-bin qubits from $\approx$1\% to $\approx$29\%, which falls only a few percent short of the previous theoretical maximum of 31\%. With further improvements to reduce photon loss in the transmission line, the Bell state measurement efficiency would only be limited by the intrinsic efficiency of the SNSPDs, which has been reported to exceed 90\% \cite{Marsili13}. Hence, the BSM efficiency can exceed 40\%, which is close to the maximum of 50\%

Bell state measurements are key ingredients for applications of quantum information processing, including linear optics quantum computing, quantum repeaters, and measurement-device-independent quantum key distribution, and our results are interesting in view of improving (or allowing) implementations. However, to take full advantage of the increased efficiency, detector dead-times need to be decreased, for instance using detector arrays \cite{Verma2014}, to allow reducing the spacing between temporal modes used to encode time-bin qubits. Shorter time-bin separations would furthermore reduce the requirement on laser stability and enable two independent sources at Alice and Bob employing commercially available lasers.

\section*{Acknowledgments}
WT, JAS, PC, AR, RV, ILM and DO thank Neil Sinclair and Vladimir Kiselyov for discussions and technical support, and acknowledge funding by Alberta Innovates Technology Futures (AITF), the National Sciences and Engineering Research Council of Canada (NSERC), the US Defense Advanced Research Projects Agency (DARPA) Quiness Program under Grant No. W31P4Q-13-1-0004, and the Killam Trusts. VBV and SWN acknowledge partial funding for detector development from the DARPA Information in a Photon (InPho) program. Part of the research was carried out at the Jet Propulsion Laboratory, California Institute of Technology, under a contract with the National Aeronautics and Space Administration.


\begin{thebibliography}{99}
\bibitem{Sanguard11}{ N. Sangouard, C. Simon, H. de Riedmatten and N. Gisin,``Quantum repeaters based on atomic ensembles and linear optics,"  Rev. Mod. Phys. \textbf{83}, 33--80 (2011).}

\bibitem{Bennett93}{C. H. Bennett, G. Brassard, C. Cr\'epeau, R. Jozsa, A. Peres and W. K. Wootters, ``Teleporting an unknown quantum state via dual classical and Einstein-Podolsky-Rosen channels,'' Phys. Rev. Lett. \textbf{70}, 1895--1899 (1993).}

\bibitem{Mattle96}{K. Mattle, H. Weinfurter, P. G. Kwiat and A. Zeilinger, ``Dense Coding in Experimental Quantum Communication," Phys. Rev. Lett. \textbf{76}, 4656--4659 (1996).}

\bibitem{Lo2012}{H.-K. Lo, M. Curty and B. Qi, ``Measurement-device-independent quantum key disitrbution," Phys. Rev. Lett. \textbf{108}, 130503 (2012).}

\bibitem{Lutkenhaus99}{N. L\"utkenhaus, J. Calsamiglia and K.-A. Suominen, ``Bell measurements for teleportation," Phys. Rev. A \textbf{59}, 3295--3300 (1999).}

\bibitem{plug&play}A. Muller, T. Herzog, B. Huttner, W. Tittel, H. Zbinden, and N. Gisin, ``Plug and play systems for quantum cryptography," Appl. Phys. Lett. \textbf{70}, 793--795 (1997).

\bibitem{Brendel}J. Brendel, N. Gisin, W. Tittel, and H. Zbinden, ``Pulsed energy-time enangled twin-photon source for quantum communication," Phys. Rev. Lett. \textbf{82}, 2594--2597 (1999).

\bibitem{Tittel2000}W. Tittel, J. Brendel, H. Zbinden, and N. Gisin, ``Quantum cryptography using entangled photons in energy-time Bell states,'' Phys. Rev. Lett. \textbf{84}, 4737--4740 (2000). 

\bibitem{Marcikic2003} I. Marcikic, H. De Riedmatten, W. Tittel, H. Zbinden, and N. Gisin, ``Long-distance teleportation of qubits at telecommunication wavelengths," Nature \textbf{421}, 509--513 (2003).

\bibitem{DeRiedmatten2005} H. De Riedmatten, I. Marcikic, J. A. W. van Houwelingen, W. Tittel, H. Zbinden, N. Gisin, ``Long-distance entanglement swapping with photons from separated sources," Phys. Rev. A \textbf{71}, 050302 (2005).

\bibitem{Jin2013} J. Jin, J. A. Slater, E. Saglamyurek, NJ. Sinclair, M. George, R. Ricken, D. Oblak, W. Sohler, and W. Tittel, ``Two-photon interference of weak coherent laser pulses recalled from separate solid-state quantum memories," Nature Comm. \textbf{4},  2386 (2013).
 
\bibitem{Rubenok13}{ A. Rubenok, J. A. Slater, P. Chan, I. Lucio-Martinez, and W. Tittel, ``Real-World Two-Photon Interference and Proof-of-Principle Quantum Key Distribution Immune to Detector Attacks," Phys. Rev. Lett. \textbf{111}, 130501 (2013).}

\bibitem{Liu13}{Y. Liu, T.-Y. Chen, L.-J. Wang, H. Liang, G.-L. Shentu, J. Wang, K. Cui, H.-L. Yin, N.-L. Liu, L. Li, X. Ma, J. S. Pelc, M. M. Fejer, C.-Z. Peng, Q. Zhang and J.-W. Pan, ``Experimental Measurement-Device-Independent Quantum Key Distribution," Phys. Rev. Lett. \textbf{111}, 130502 (2013).}

\bibitem{Dixon09}{A. R. Dixon and J. F. Dynes and Z. L. Yuan and A. W. Sharpe and A. J. Bennet and A. J. Shields, ``Ultrashort dead time of photon-counting InGaAs avalanche photodiodes,'' Appl. Phys. Lett. \textbf{94}, 231113 (2009).}

\bibitem{Zhang09}{J. Zhang, R. Thew, C. Barreiro and H. Zbinden, ``Practical fast gate rate InGaAs/InP single-photon avalanche photodiodes,'' Appl. Phys. Lett. \textbf{95}, 091103 (2009).}



\bibitem{Zhang08}{J. Zhang, R. Thew, J.-D. Gautier, N. Gisin, H. Zbinden, ``Comprehensive Characterization of InGaAs/InP Avalanche Photodiodes at 1550nm with an Active Quenching ASIC," J. Quantum Electron. \textbf{45}, 792--799 (2009).}

\bibitem{Houwelingen06}J. A. W. van Houwelingen, N. Brunner, A. Beveratos, H. Zbinden, and N. Gisin, ``Quantum Teleportation with a Three-Bell-State Analyzer," Phys. Rev. Lett. \textbf{96}, 130502 (2006).

\bibitem{Natarajan2012a}{C. M. Natarajan, M. G. Tanner, and R. H. Hadfield, `` Superconducting nanowire
single-photon detectors: physics and applications,'' Supercond. Sci. Technol. \textbf{25}, 063001 (2012).}

\bibitem{Kerman2006a}{A. J. Kerman, E. A. Dauler, W. E. Keicher, J. K. W. Yang, K. K. Berggren, G. GolÕtsman and B. Voronov, ``Kinetic-inductance-limited reset time of superconducting nanowire photon counters,'' Appl. Phys. Lett. \textbf{88}, 111116 (2006).}

\bibitem{Robinson2006a}B. S. Robinson, A. J. Kerman, E. A. Dauler, R. J. Barron, D. O. Caplan, M. L. Stevens, J. J. Carney, S. A. Hamilton, J. K. Yang, and K. K. Berggren,``781-Mbit/s photon-counting optical communications using a super-conducting nanowire detector,'' Opt. Lett. \textbf{31}, 444--446 (2006).

\bibitem{Marsili13}{F. Marsili, V. B. Verma, J. A. Stern, S. Harrington, A. E. Lita,  T. Gerrits, I. Vayshenker, B. Baek, M. D. Shaw, R. P. Mirin and S. W. Nam, ``Detecting single infrared photons with 93\% system efficiency," Nat. Photon. \textbf{7}, 210-214  (2013).} 

\bibitem{Verma12}{V. B. Verma, F. Marsili, S. Harrington, A. E. Lita, R. P. Mirin, and S. W. Nam, ``A three-dimensional, polarization-insensitive superconducting nanowire avalanche photodetector,'' Appl. Phys. Lett. \textbf{101}, 251114 (2012).}

\bibitem{Yang2007a}{J. K. W. Yang, A. J. Kerman, E. A. Dauler, V. Anant, K. M. Rosfjord and K. K. Berggren, 
``Modeling the Electrical and Thermal Response of Superconducting Nanowire Single-Photon Detectors,'' IEEE Trans. on  Appl. Supercond. \textbf{17}, 581--585 (2007).}

\bibitem{Wang2013}{X.-B. Wang, ``Three-intensity decoy state method for device independent quantum key distribution with basis dependent errors,'' Phys. Rev. A \textbf{87}, 012320 (2013).}


\bibitem{Verma2014}{V. B. Verma, R. Horansky, F. Marsili, J. A. Stern, M. D. Shaw, A. E. Lita, R. P. Mirin, S. W. Nam, ''A four-pixel single-photon pulse-position array fabricated from WSi superconducting nanowire single-photon detectors,'' Appl. Phys. Lett. \textbf{104}, 051115 (2014)}




\end{thebibliography}
\end{document}